\documentclass[aps,prl,twocolumn,showpacs,amsmath,amssymb,floatfix,superscriptaddress]{revtex4}
\usepackage{graphicx}
\newcommand{\Levy}{L\'evy } 

\begin{document}

\title{Density Profiles in Open Superdiffusive Systems}
\date{\today}

\author{Stefano Lepri}
\affiliation{Consiglio Nazionale
delle Ricerche, Istituto dei Sistemi Complessi, 
via Madonna del Piano 10, I-50019 Sesto Fiorentino, Italy}

\author{Antonio Politi}
\affiliation{Consiglio Nazionale
delle Ricerche, Istituto dei Sistemi Complessi, 
via Madonna del Piano 10, I-50019 Sesto Fiorentino, Italy}

\begin{abstract}
We numerically solve a discretized model of \Levy random walks on a finite
one-dimensional domain in the presence of sources and with a reflection
coefficient $r$. At the domain boundaries, the steady-state density profile
is non-analytic. The meniscus exponent $\mu$, introduced to characterize this
singular behavior, uniquely identifies the whole profile. Numerical data
suggest that $\mu =\alpha/2 + r(\alpha/2-1)$, where $\alpha$ is the \Levy exponent
of the step-length distribution. As an application, we show that this model
reproduces the temperature profiles obtained for a chain of oscillators
displaying anomalous heat conduction. Remarkably, the
case of free-boundary conditions in the chain correspond to a \Levy walk with
negative reflection coefficient.   
\end{abstract}


\maketitle
Anomalous diffusion and transport are ubiquitous
phenomena that occur in many complex systems, ranging from molecular motion to
human mobility \cite{AnotransBook08}. In the framework of random walk theory
much progress has been made in the last years in the characterization of
superdiffusive motion in infinite domains  \cite{Metzler2004,AnotransBook08}.
Extension to the case of finite systems is relatively less developed
\cite{Drysdale1998,Buldyrev2001}. Typically, only the simple case of absorbing
Boundary Conditions (BC) without sources has been treated
\cite{vanMilligen08}. An even less studied problem is that of \textit{finite and
open} superdiffusive systems where particles and energy are steadily exchanged
with some external reservoirs \cite{Denisov03}. 
This setup describes a variety of systems with anomalous transport
in the off-equilibrium regime. 
For many-particle systems this is exemplified by anomalous heat conduction
in low-dimensional
lattices \cite{LLP03,Lepri10}, as well as by enhanced thermodiffusion
effects \cite{Saito2010}. Moreover, there exist at least two physical setups
where the theoretical predictions can be experimentally tested: light
superdiffusion \cite{Barthelemy2008,Barthelemy2010} and anomalous heat
conduction in individual nanotubes \cite{Chang08}. 

In the present Letter we study the spatial distribution of
the relevant field (particle density, temperature, light intensity,
depending on the context) in the presence of an external gradient. By
numerically investigating a model of particles performing \Levy flights
or walks, we find that the shape of the profile is nonlinear in the
thermodynamic/continuum limit. This is due to the long-range correlations which
affect the transport far away from the boundaries. A comparison with the
temperature profiles obtained in a model of heat conductivity indicates that
the shape is fully determined by the exponent of the underlying \Levy process
and by the BC. However, the connection between the (free/fixed) BC of the 
conduction problem and those of the diffusive process (reflectivity)
is not trivial. We indeed find that free BC correspond
to a negative reflectivity.

\begin{figure}[h!]
\includegraphics[width=0.4\textwidth,clip]{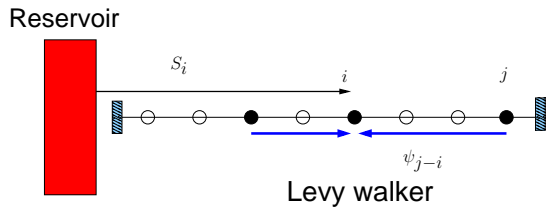}
\caption{(Color online) Scheme of the model: an ensemble
of \Levy walkers on a finite one-dimensional lattice
interacting with one external reservoir that 
inject particles with a prescribed constant rate $S_i$
at each site. 
}
\label{fig1}
\end{figure}

\textit{The model -} Let $i$ denote the position of a discrete-time  
random walker on a finite one-dimensional lattice ($1\le i\le N)$. 
In between consecutive scattering events, the particle either jumps 
instantaneously (\Levy flight - LF) or moves with unit velocity 
(\Levy walk - LW \cite{Blumen1989}) over a distance of $m$ sites, 
that is randomly selected according to the step-length distribution 
\begin{equation}
\label{eq:dist}
\psi_m \; = \; \frac{q}{|m|^{1+\alpha}},\quad \psi_0=0.
\end{equation}  
Here $\alpha$ is the \Levy exponent and $q$ is a normalization constant.
We limit ourselves to the case $1\leq \alpha \le 2$, 
where the averages are finite but the variance not. 
We formulate the problem by introducing the vector ${\bf W}\equiv\{ W_i(t)\}$,
where $W_i$ is the probability for the walker to undergo a scattering event at
site $i$ and time $t$. It satisfies a master equation, which, for LFs, writes
\begin{equation}
{\bf W}(t+1) = {\bf Q}{\bf W}(t) +{\bf S} ,
\label{eq:sme}
\end{equation}
where ${\bf S}$ accounts for the particles steadily injected from external
resorvoirs (see Fig.~\ref{fig1} for a pictorial representation); ${\bf Q}$ is a
matrix describing the probability of paths connecting pair of sites. In the
simple case of absorbing BC, it is readily seen
that $Q_{ji}$ is equal to the probability $\psi_{j-i}$ of a direct flight, as
from Eq.~(\ref{eq:dist}) (see  Fig.~\ref{fig1}). In the LW case, the ${\bf W}$
components in the r.h.s. must be estimated at different times (depending on the
length of the path followed from $j$ to $i$). Since, the stationary solution is
the same in both setups, this difference is immaterial, and we thereby refer to
LFs, since Eq.~(\ref{eq:sme}) is easier to solve iteratively.

In the LF case, $W_i$ is equal to the density $P_i$ of particle at site $i$.
This is no longer true for the LW, as $P_i$ includes those particles that 
are crossing during a ballistic step. In the simple case of absorbing BC, 
one can write
\begin{equation}
P_i = W_i + \sum_{j=1}^N R_{|i-j|}W_j + I_i
\label{prob}
\end{equation}
where $R_i=\sum_{j > i} \psi_j$ and $I_i=\sum_{j > i} S_j$ 
account for the particles which, having started their walk 
respectively inside or outside the domain,
transit at the $i$th site.

The source term is fixed by assuming that the reservoir is a semi-infinite
lattice, homogeneously filled by \Levy walkers of the same type as those
residing in the domain. This amounts to defining $S_m = s\, m^{-\alpha}$, where
$s$ measures the density of particles and $m$ the distance from the
reservoir. It is easy to verify that in the presence of two identical 
reservoirs at the lattice ends, the density is constant
(for any $N$), showing that our definition satisfies a kind of
``zeroth principle"  as it should.

\begin{figure}[h!]
\includegraphics[width=0.4\textwidth,clip]{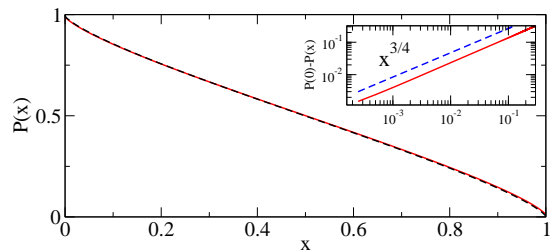}
\caption{(Color online) Density profiles 
as a function of the variable $x=i/N$ for 
absorbing BC and $\alpha=3/2$. Solid line: solution of
the  
Eq.~(\ref{eq:sme}) with $N=4000$. Dashed line: solution of the
FDE discretized on a mesh of 
$M=800$ points; the latter curve is
rescaled by a suitable constant. The inset
is a magnification of the leftmost region.
}
\label{fig2}
\end{figure}

In the nonequilibrium case, it is not necessary to deal with two reservoirs. The
linearity of the problem teaches us that it is sufficient to study the case of a
single reservoir, that we assume to be in $i=0$: the effect of, say, a second
one on the opposite side can be accounted for by a suitable linear
transformation. For values of $N$ large enough, the steady-state density 
depends on $i$ and $N$ through the combined variable $x=i/N$, $P_i=P(x)$. 
As seen in Fig.~\ref{fig2}, $P$ vanishes for $x \to 1$ because on that side 
the absorbing boundary is not accompanied by an incoming flux of particles. Besides, we  have also verified
that $W_i \propto P_i$ in agreement with the expectation that LFs and LWs should
have by the same stationary properties.

\textit{Fractional Laplacian-}  
For a more accurate consistency check, we now compare the solution of our model
with that of the stationary Fractional Diffusion Equation (FDE)
$D^\alpha_x P=-\sigma(x)$
on the interval $0\le x \le 1$ (see e.g. Ref.~\cite{Zoia07} 
and references therein for a precise 
definition of the operator $D^\alpha_x$). The source term $\sigma(x)$ 
must be chosen so as to describe the effect of the external reservoirs.  
A condition to be fulfilled is that two identical reservoirs yield
a homogeneous state $P(x)=const.$\,. A straightforward calculation 
using the integral definition of the operator \cite{Zoia07} shows 
that this happens for $\sigma(x) = \sigma_{eq}(x) 
\equiv x^{-\alpha}+(1-x)^{-\alpha}$ 
(we, henceforth, ignore irrelevant proportionality constants) 
\footnote{The FDE is solved by  discretizing the fractional Laplacian on a mesh
of $M$ equally-spaced sites as given in Ref.~\cite{Zoia07}. A particular care
is  required to correctly treat the discretization $\sigma_i$ of the  source at
$i=1,M$. Empirically, we adjusted $\sigma_{1,M}$ to ensure that the equilibrium
solution $P=const.$ is accurately  reproduced for $\sigma=\sigma_{eq}$.
}.
It is thus natural to associate $\sigma(x) = x^{-\alpha}$ to the 
nonequilibrium case with a single source in $x=0$.
The numerical solution of the FDE
is reported  in Fig.~\ref{fig2} (dashed line). It overlaps almost 
perfectly with the stationary solution of the discrete model, 
thereby confirming the correctness of our choice. Moreover, this 
shows that long-ranged sources are needed to reproduce the 
superdiffusive profiles in the continuum limit.

\textit{Boundary singularity -} A distinctive feature of the profile 
is that it is not analytic at the boundaries. Indeed, the data for 
$x\to 0$ are well fitted by
\begin{equation}
P(x) \;=\; P(0) + Cx^\mu
\end{equation}
(the same behavior occurs for $x\to 1$, as the profiles are
symmetric). In view of the similarity with the shape of the liquid
surface close to a wall, we metaphorically term $\mu$ as the 
\textit{meniscus exponent}. The density
drop is akin to the well-known Kapitza resistance, which has been previously
discussed for one-dimensional systems \cite{Aoki01}. However, the
nonanalytic behavior is peculiar of anomalous kinetics, 
as opposed to the familiar linear shape in standard diffusion.   
For the above discussed case of absorbing boundaries, we find that
$\mu\approx \alpha/2$. This value is consistent with the  
singular behavior of the eigenfunctions of the fractional Laplacian 
\cite{Zoia07}.

\textit{Reflecting boundaries -} We now generalize the discrete model, by
assuming that any time a particle reaches the boundary, it is reflected with
probability $r$ ($0 \le r \le 1$). This extension is of central interest
for real problems, e.g. in anomalous diffusion of light \cite{Bertolotti2010}, 
but we have to pay the price of a more complicate mathematical treatment
to account for multiple reflections. In fact,
the transition probability $Q_{ji}$ from the site $j$ to site $i$ is the sum of infinitely many terms
that correspond to paths undergoing multiple reflections. By calling $B_l$ and
$B_r$ the left and right boundaries,
respectively, and viewing any path as a sequence of symbols that identify
the starting and final site, as well as all reflection points, the four
families that contribute to $Q_{ji}$ can be encoded as (for $1<j<i<N-1$):
$j(B_rB_l)^mi$, $jB_l(B_rB_l)^mi$, $jB_r(B_lB_r)^mi$, and $j(B_lB_r)^{m+1}i$ with
$m=0,1,\ldots$. Symmetric relations hold for $j>i$, while special expressions
must be invoked for the degenerate cases ($i,j=1,N-1$, $i=j$). The weight of
a path of length $\ell$ is then given by $r^k \psi_\ell$, where $k$ is the number
of reflections ($k=2m+1$ in the second of the four families).
By summing up all contributions for the same pairs of sites, we obtain
$Q_{ji}$. A similar treatment must be applied to the source term to take into
account that each particle entering from either side may be trapped
for an arbitrarily long time. Another complication is that for LWs, the 
expression for $P(x)$ requires including all paths arising from multiple 
reflections. Notice however, that all such intricacies
do not change the rank of the matrix $\bf Q$.

\begin{figure}[ht!]
\includegraphics[width=0.38\textwidth,clip]{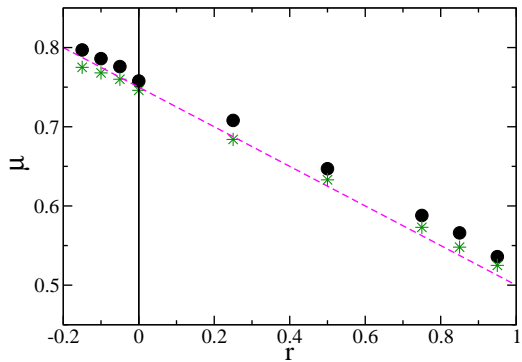}
\caption{(Color online) Dependence of the meniscus exponent $\mu$ on the 
reflection probability $r$ for $\alpha=3/2$. Full circles and stars 
are measures from fitting of $P(x)$ and $W(x)$, respectively (see text). 
The dashed line is given by the formula (\ref{eq:conj}). 
}
\label{fig5}
\end{figure}

Unexpectedly, we found that the shape of the profile depends on $r$ for 
fixed $\alpha$.
This is entirely new with respect to normal diffusion, when the profile is 
linear irrespective of the boundary conditions. The meniscus exponent 
is a  convenient tool to parametrize this functional dependence on $r$.
The estimated numerical values of $\mu$ for $\alpha = 3/2$ are reported in
Fig.~\ref{fig5}. Once more, we found that 
$W_i \propto P_i$ and that, accordingly, both profiles give compatible 
values of $\mu$ (see symbols in Fig.~\ref{fig5}). By assuming a linear dependence of $\mu$ on  both
$r$, and $\alpha$, we conjecture
\begin{equation}
\label{eq:conj}
\mu \;=\;\frac{\alpha}{2} + r \left(\frac{\alpha}{2}-1\right)\,.
\end{equation}
This expression is consistent with the $\mu=\alpha/2$ value found 
above for $r=0$. Moreover, for $\alpha=2$ (normal diffusion) 
it yields $\mu=1$, as it should.
The discrepancy between formula (\ref{eq:conj}) and the numerical values
is everywhere compatible with the finite-size corrections (data has been
obtained with $N=8000$). 

\textit{Application to the heat conduction problem-} Temperature profiles in
one-dimensional systems displaying  anomalous energy transport \cite{LLP03}
represent a good testing  ground of the above approach. In fact, energy
perturbations do spread  like in a \Levy process \cite{Cipriani05}. Also, 
for an
infinite chain of harmonic oscillators with energy and momentum conserving
noise \cite{BBO06}, it was proven that the phonon  Boltzmann equation reduces to
a FDE  with $\alpha=3/2$ \cite{Jara09}. 
We thus expect model (\ref{eq:sme}) to apply to a finite chain of
oscillators, (i) coupled with two Langevin heat baths (with a damping constant
$\lambda$), and (ii) with random collisions that exchange the velocities of
neighboring particles with a rate $\gamma$ \cite{DLLP08}.
This model has the advantage of allowing for an exact solution of the 
associated Fokker-Planck equation \cite{Lepri2009}.

\begin{figure}[!ht]
\includegraphics[width=0.4\textwidth,clip]{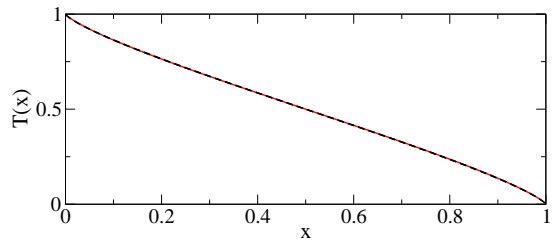}
\caption{(Color online) Temperature profile of the 
oscillator chain with conservative noise 
with free boundary condition and  $\lambda=\gamma=1$
(solid line) and solution of the master equation
with reflection coefficient $r=-0.1$ (dashed line).
}
\label{fig4}
\end{figure}

In Fig.~\ref{fig4} we compare the temperature profile $T(x)$ (suitably shifted
and rescaled) of the heat-conduction model with free BC and the solution of our
discrete \Levy model with a reflection coefficient $r=-0.1$. Since they are
essentially indistinguishable, we can conclude that the \Levy interpretation does
not only allow to explain the anomalous scaling of heat conductivity
\cite{Cipriani05}, but also the peculiar shape of $T(x)$. The weird (negative)
value of $r$ requires a specific discussion. Since ${\bf Q}$ is the sum of 
terms proportional to $r^k$ ($k$ being 
the number of reflections from the boundaries),  even and odd $k$s
yield, for $r<0$, positive and negative contributions, respectively. 
Thus, we let ${\bf Q}= {\bf Q}_e-{\bf Q}_o$. The presence of an unphysical 
negative term can be avoided by introducing two families of 
walkers (${\bf W}^+$ and ${\bf W}^-$) and interpreting the reflection 
as a change of family. Under these assumptions, the evolution 
equations read
\begin{equation}
{\bf W}^\pm(t+1) = {\bf Q}_e{\bf W}^\pm(t) + {\bf Q}_o{\bf W}^\mp(t) +
{\bf S}^\pm \, ,
\end{equation}
where all terms are positive as they should. Under the further assumption that
${\bf S}^+={\bf S}$ and ${\bf S}^-=0$, we can easily see that
${\bf W} = {\bf W}^+-{\bf W}^-$ satisfies the original Eq.~(\ref{eq:sme}). 
In other words, the relevant quantity to look at is the difference between the
densities of the two different families. Why it is necessary to invoke the
presence of such two families and what is their physical meaning in the context
of heat conductivity is an open problem.

In the case of a chain with fixed BC, the temperature profile $T(x)$ 
can be computed analytically \cite{Lepri2009} and it is thereby found 
that $\mu=1/2$. By inserting this value in 
Eq.~(\ref{eq:conj}) and recalling that $\alpha=3/2$, we find that $r=1$, 
i.e. the fixed-BC $T(x)$ corresponds to the case of perfectly reflecting 
barriers. Unfortunately, this (physically reasonable) result could not
be tested quantitatively. Indeed, it turns out that finite-size
corrections become increasingly important upon increasing $r$, and for 
$r$ close to 1, it is practically impossible to achieve convergence
to the steady-state. 

\begin{figure}[h!]
\includegraphics[width=0.4\textwidth,clip]{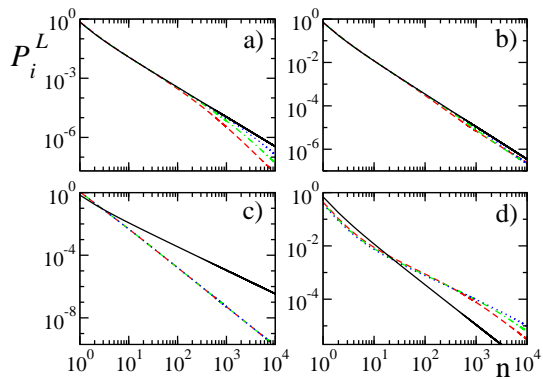}
\caption{(Color online) Distributions of left moving 
particles $P_i^L(n)$ (broken lines) for $i=N/2$  and 
(a,b) and $i=N$ (c,d). Dashed, dotted-dashed, and dotted curves 
correspond to $N=1000$, 4000 and 16000, respectively; the solid 
line is $P_{eq}(n)$; see text for details. 
All curves are rescaled to span unit area.}
\label{fig3}
\end{figure}

\textit{Local equilibrium-} In the context of LWs, we can go beyond the
determination of the shape of $P$ and check the existence of local equilibrium.
To this aim, we decompose $P_i$ as $\sum_n [P_i^R(n)+P_i^L(n)]$ 
where $P_i^{L,R}(n)$ is the
probability that a walker on site $i$ and moving respectively 
to the left or right will undergo the next scattering event after $n$ 
time steps. Then, we compare $P_i^{L,R}(n)$ with its 
equilibrium value $P_{eq}(n)$, obtained from the stationary solution of 
Eq.~(\ref{eq:sme}) with two equal sources on both sides
[$P_{eq}(n)$ is independent of $i$ and of $L,R$].
Unsurprisingly, $P_i^{L,R}(n)\simeq P_{eq}(n)$ close 
to the left boundary. 
Moving away from the reservoir, the densities $P_i^L(n)$ 
start to deviate from $P_{eq}(n)$ (see respectively the broken and solid
lines in Fig.~\ref{fig3}). 
The $P_i^R(n)$ (not reported) display similar, 
though somehow smaller, deviations. Fig.~\ref{fig3}a and b refer to the middle 
of the system and $r=0$, $r=0.5$, respectively. Deviations decrease upon
increasing the system sizes and are larger in the case of absorbing BC, as the
particles have less time to ``thermalize". 
Fig.~\ref{fig3} c and d refer to the rightmost site, again for $r=0$ and $r=0.5$. In this case there is no evidence of a
convergence towards the equilibrium distribution, but this result is not
earthshaking, as the density of particles decreases for increasing $N$. In other
words, the deviations concern an asymptotically vanishing fraction of particles.

\textit{Conclusions -} In this Letter we have determined the density profiles in
the case of superdiffusive motion in an open, finite domain for arbitrary 
reflectivity $r$. Unexpectedly, their shape depends on both $\alpha$ and $r$
through the meniscus exponent $\mu$, Eq.~(\ref{eq:conj}). This is a
manifestation of the long-range effects induced by superdiffusive behavior. The
prediction of the \Levy model accounts quantitatively  for the shape of the
temperature profile of an interacting oscillator chain displaying anomalous heat
conductivity. It remains to clarify the relationship between the type of BC in
the thermal problem with that in the diffusive one, that appears to be far from
trivial. Another open problem is the rigorous proof of the
conjecture (\ref{eq:conj}). We finally hope that the shape can be observed in
some optical experiment.

We thank J. Bertolotti for fruitful discussions. This work is part of the
PRIN project 2008Y4W3CY.

\bibliography{levy,heat}

\begin{thebibliography}{21}
\expandafter\ifx\csname natexlab\endcsname\relax\def\natexlab#1{#1}\fi
\expandafter\ifx\csname bibnamefont\endcsname\relax
  \def\bibnamefont#1{#1}\fi
\expandafter\ifx\csname bibfnamefont\endcsname\relax
  \def\bibfnamefont#1{#1}\fi
\expandafter\ifx\csname citenamefont\endcsname\relax
  \def\citenamefont#1{#1}\fi
\expandafter\ifx\csname url\endcsname\relax
  \def\url#1{\texttt{#1}}\fi
\expandafter\ifx\csname urlprefix\endcsname\relax\def\urlprefix{URL }\fi
\providecommand{\bibinfo}[2]{#2}
\providecommand{\eprint}[2][]{\url{#2}}

\bibitem[{\citenamefont{Klages et~al.}(2008)\citenamefont{Klages, Radons, and
  Sokolov}}]{AnotransBook08}
\bibinfo{editor}{\bibfnamefont{R.}~\bibnamefont{Klages}},
  \bibinfo{editor}{\bibfnamefont{G.}~\bibnamefont{Radons}}, \bibnamefont{and}
  \bibinfo{editor}{\bibfnamefont{I.~M.} \bibnamefont{Sokolov}}, eds.,
  \emph{\bibinfo{title}{Anomalous Transport: Foundations and Applications}}
  (\bibinfo{publisher}{Wiley-VCH Verlag, Weinheim}, \bibinfo{year}{2008}).

\bibitem[{\citenamefont{Metzler and Klafter}(2004)}]{Metzler2004}
\bibinfo{author}{\bibfnamefont{R.}~\bibnamefont{Metzler}} \bibnamefont{and}
  \bibinfo{author}{\bibfnamefont{J.}~\bibnamefont{Klafter}},
  \bibinfo{journal}{J. Phys A: Math. Gen.} \textbf{\bibinfo{volume}{37}},
  \bibinfo{pages}{R161} (\bibinfo{year}{2004}).

\bibitem[{\citenamefont{Buldyrev et~al.}(2001)\citenamefont{Buldyrev, Havlin,
  Kazakov, da~Luz, Raposo, Stanley, and Viswanathan}}]{Buldyrev2001}
\bibinfo{author}{\bibfnamefont{S.~V.} \bibnamefont{Buldyrev}},
  \bibinfo{author}{\bibfnamefont{S.}~\bibnamefont{Havlin}},
  \bibinfo{author}{\bibfnamefont{A.~Y.} \bibnamefont{Kazakov}},
  \bibinfo{author}{\bibfnamefont{M.~G.~E.} \bibnamefont{da~Luz}},
  \bibinfo{author}{\bibfnamefont{E.~P.} \bibnamefont{Raposo}},
  \bibinfo{author}{\bibfnamefont{H.~E.} \bibnamefont{Stanley}},
  \bibnamefont{and} \bibinfo{author}{\bibfnamefont{G.~M.}
  \bibnamefont{Viswanathan}}, \bibinfo{journal}{Phys. Rev. E}
  \textbf{\bibinfo{volume}{64}}, \bibinfo{pages}{041108}
  (\bibinfo{year}{2001}).

\bibitem[{\citenamefont{Drysdale and Robinson}(1998)}]{Drysdale1998}
\bibinfo{author}{\bibfnamefont{P.~M.} \bibnamefont{Drysdale}} \bibnamefont{and}
  \bibinfo{author}{\bibfnamefont{P.~A.} \bibnamefont{Robinson}},
  \bibinfo{journal}{Phys. Rev. E} \textbf{\bibinfo{volume}{58}},
  \bibinfo{pages}{5382} (\bibinfo{year}{1998}).

\bibitem[{\citenamefont{van Milligen et~al.}(2008)\citenamefont{van Milligen,
  Calvo, and S\'anchez}}]{vanMilligen08}
For continuous-time random walks, general BC have been recently considered in
\bibinfo{author}{\bibfnamefont{B.~P.} \bibnamefont{van Milligen}},
  \bibinfo{author}{\bibfnamefont{I.}~\bibnamefont{Calvo}}, \bibnamefont{and}
  \bibinfo{author}{\bibfnamefont{R.}~\bibnamefont{S\'anchez}},
  \bibinfo{journal}{J. Phys. A: Math. Theor.} \textbf{\bibinfo{volume}{41}},
  \bibinfo{pages}{215004} (\bibinfo{year}{2008}).

\bibitem[{\citenamefont{Denisov et~al.}(2003)\citenamefont{Denisov, Klafter,
  and Urbakh}}]{Denisov03}
\bibinfo{author}{\bibfnamefont{S.}~\bibnamefont{Denisov}},
  \bibinfo{author}{\bibfnamefont{J.}~\bibnamefont{Klafter}}, \bibnamefont{and}
  \bibinfo{author}{\bibfnamefont{M.}~\bibnamefont{Urbakh}},
  \bibinfo{journal}{Phys. Rev. Lett.} \textbf{\bibinfo{volume}{91}},
  \bibinfo{pages}{194301} (\bibinfo{year}{2003}).

\bibitem[{\citenamefont{Lepri et~al.}(2010)\citenamefont{Lepri,
  Mej{\'\i}a-Monasterio, and Politi}}]{Lepri10}
\bibinfo{author}{\bibfnamefont{S.}~\bibnamefont{Lepri}},
  \bibinfo{author}{\bibfnamefont{C.}~\bibnamefont{Mej{\'\i}a-Monasterio}},
  \bibnamefont{and} \bibinfo{author}{\bibfnamefont{A.}~\bibnamefont{Politi}},
  \bibinfo{journal}{J. Phys. A: Math. Theor.} \textbf{\bibinfo{volume}{43}},
  \bibinfo{pages}{065002} (\bibinfo{year}{2010}).

\bibitem[{\citenamefont{Lepri et~al.}(2003)\citenamefont{Lepri, Livi, and
  Politi}}]{LLP03}
\bibinfo{author}{\bibfnamefont{S.}~\bibnamefont{Lepri}},
  \bibinfo{author}{\bibfnamefont{R.}~\bibnamefont{Livi}}, \bibnamefont{and}
  \bibinfo{author}{\bibfnamefont{A.}~\bibnamefont{Politi}},
  \bibinfo{journal}{Phys. Rep.} \textbf{\bibinfo{volume}{377}},
  \bibinfo{pages}{1} (\bibinfo{year}{2003}).

\bibitem[{\citenamefont{Saito et~al.}(2010)\citenamefont{Saito, Benenti, and
  Casati}}]{Saito2010}
\bibinfo{author}{\bibfnamefont{K.}~\bibnamefont{Saito}},
  \bibinfo{author}{\bibfnamefont{G.}~\bibnamefont{Benenti}}, \bibnamefont{and}
  \bibinfo{author}{\bibfnamefont{G.}~\bibnamefont{Casati}},
  \bibinfo{journal}{Chem. Phys.} \textbf{\bibinfo{volume}{375}},
  \bibinfo{pages}{508} (\bibinfo{year}{2010}).

\bibitem[{\citenamefont{Barthelemy et~al.}(2008)\citenamefont{Barthelemy,
  Bertolotti, and Wiersma}}]{Barthelemy2008}
\bibinfo{author}{\bibfnamefont{P.}~\bibnamefont{Barthelemy}},
  \bibinfo{author}{\bibfnamefont{J.}~\bibnamefont{Bertolotti}},
  \bibnamefont{and} \bibinfo{author}{\bibfnamefont{D.}~\bibnamefont{Wiersma}},
  \bibinfo{journal}{Nature} \textbf{\bibinfo{volume}{453}},
  \bibinfo{pages}{495} (\bibinfo{year}{2008}).

\bibitem[{\citenamefont{Barthelemy et~al.}(2010)\citenamefont{Barthelemy,
  Bertolotti, Vynck, Lepri, and Wiersma}}]{Barthelemy2010}
\bibinfo{author}{\bibfnamefont{P.}~\bibnamefont{Barthelemy}},
  \bibinfo{author}{\bibfnamefont{J.}~\bibnamefont{Bertolotti}},
  \bibinfo{author}{\bibfnamefont{K.}~\bibnamefont{Vynck}},
  \bibinfo{author}{\bibfnamefont{S.}~\bibnamefont{Lepri}}, \bibnamefont{and}
  \bibinfo{author}{\bibfnamefont{D.~S.} \bibnamefont{Wiersma}},
  \bibinfo{journal}{Phys. Rev. E} \textbf{\bibinfo{volume}{82}},
  \bibinfo{pages}{011101} (\bibinfo{year}{2010}).

\bibitem[{\citenamefont{Chang et~al.}(2008)\citenamefont{Chang, Okawa, Garcia,
  Majumdar, and Zettl}}]{Chang08}
\bibinfo{author}{\bibfnamefont{C.~W.} \bibnamefont{Chang}},
  \bibinfo{author}{\bibfnamefont{D.}~\bibnamefont{Okawa}},
  \bibinfo{author}{\bibfnamefont{H.}~\bibnamefont{Garcia}},
  \bibinfo{author}{\bibfnamefont{A.}~\bibnamefont{Majumdar}}, \bibnamefont{and}
  \bibinfo{author}{\bibfnamefont{A.}~\bibnamefont{Zettl}},
  \bibinfo{journal}{Phys. Rev. Lett.} \textbf{\bibinfo{volume}{101}},
  \bibinfo{pages}{075903} (\bibinfo{year}{2008}).

\bibitem[{\citenamefont{Blumen et~al.}(1989)\citenamefont{Blumen, Zumofen, and
  Klafter}}]{Blumen1989}
\bibinfo{author}{\bibfnamefont{A.}~\bibnamefont{Blumen}},
  \bibinfo{author}{\bibfnamefont{G.}~\bibnamefont{Zumofen}}, \bibnamefont{and}
  \bibinfo{author}{\bibfnamefont{J.}~\bibnamefont{Klafter}},
  \bibinfo{journal}{Phys. Rev. A} \textbf{\bibinfo{volume}{40}},
  \bibinfo{pages}{3964} (\bibinfo{year}{1989}).

\bibitem[{\citenamefont{Zoia et~al.}(2007)\citenamefont{Zoia, Rosso, and
  Kardar}}]{Zoia07}
\bibinfo{author}{\bibfnamefont{A.}~\bibnamefont{Zoia}},
  \bibinfo{author}{\bibfnamefont{A.}~\bibnamefont{Rosso}}, \bibnamefont{and}
  \bibinfo{author}{\bibfnamefont{M.}~\bibnamefont{Kardar}},
  \bibinfo{journal}{Phys. Rev. E} \textbf{\bibinfo{volume}{76}},
  \bibinfo{pages}{21116} (\bibinfo{year}{2007}).

\bibitem[{\citenamefont{Aoki and Kusnezov}(2001)}]{Aoki01}
\bibinfo{author}{\bibfnamefont{K.}~\bibnamefont{Aoki}} \bibnamefont{and}
  \bibinfo{author}{\bibfnamefont{D.}~\bibnamefont{Kusnezov}},
  \bibinfo{journal}{Phys. Rev. Lett.} \textbf{\bibinfo{volume}{86}},
  \bibinfo{pages}{4029} (\bibinfo{year}{2001}).

\bibitem[{\citenamefont{Bertolotti et~al.}(2010)\citenamefont{Bertolotti,
  Vynck, and Wiersma}}]{Bertolotti2010}
\bibinfo{author}{\bibfnamefont{J.}~\bibnamefont{Bertolotti}},
  \bibinfo{author}{\bibfnamefont{K.}~\bibnamefont{Vynck}}, \bibnamefont{and}
  \bibinfo{author}{\bibfnamefont{D.~S.} \bibnamefont{Wiersma}},
  \bibinfo{journal}{Phys. Rev. Lett.} \textbf{\bibinfo{volume}{105}},
  \bibinfo{pages}{163902} (\bibinfo{year}{2010}).

\bibitem[{\citenamefont{Cipriani et~al.}(2005)\citenamefont{Cipriani, Denisov,
  and Politi}}]{Cipriani05}
\bibinfo{author}{\bibfnamefont{P.}~\bibnamefont{Cipriani}},
  \bibinfo{author}{\bibfnamefont{S.}~\bibnamefont{Denisov}}, \bibnamefont{and}
  \bibinfo{author}{\bibfnamefont{A.}~\bibnamefont{Politi}},
  \bibinfo{journal}{Phys. Rev. Lett.} \textbf{\bibinfo{volume}{94}},
  \bibinfo{eid}{244301} (\bibinfo{year}{2005}).

\bibitem[{\citenamefont{Basile et~al.}(2006)\citenamefont{Basile, Bernardin,
  and Olla}}]{BBO06}
\bibinfo{author}{\bibfnamefont{G.}~\bibnamefont{Basile}},
  \bibinfo{author}{\bibfnamefont{C.}~\bibnamefont{Bernardin}},
  \bibnamefont{and} \bibinfo{author}{\bibfnamefont{S.}~\bibnamefont{Olla}},
  \bibinfo{journal}{Phys. Rev. Lett.} \textbf{\bibinfo{volume}{96}},
  \bibinfo{pages}{204303} (\bibinfo{year}{2006}).

\bibitem[{\citenamefont{Jara et~al.}(2009)\citenamefont{Jara, Komorowski, and
  Olla}}]{Jara09}
\bibinfo{author}{\bibfnamefont{M.}~\bibnamefont{Jara}},
  \bibinfo{author}{\bibfnamefont{T.}~\bibnamefont{Komorowski}},
  \bibnamefont{and} \bibinfo{author}{\bibfnamefont{S.}~\bibnamefont{Olla}},
  \bibinfo{journal}{The Annals of Applied Probability}
  \textbf{\bibinfo{volume}{19}}, \bibinfo{pages}{2270} (\bibinfo{year}{2009}).

\bibitem[{\citenamefont{Delfini et~al.}(2008)\citenamefont{Delfini, Lepri,
  Livi, and Politi}}]{DLLP08}
\bibinfo{author}{\bibfnamefont{L.}~\bibnamefont{Delfini}},
  \bibinfo{author}{\bibfnamefont{S.}~\bibnamefont{Lepri}},
  \bibinfo{author}{\bibfnamefont{R.}~\bibnamefont{Livi}}, \bibnamefont{and}
  \bibinfo{author}{\bibfnamefont{A.}~\bibnamefont{Politi}},
  \bibinfo{journal}{Phys. Rev. Lett.} \textbf{\bibinfo{volume}{101}},
  \bibinfo{eid}{120604} (\bibinfo{year}{2008}).

\bibitem[{\citenamefont{Lepri et~al.}(2009)\citenamefont{Lepri,
  Mej{\'\i}a-Monasterio, and Politi}}]{Lepri2009}
\bibinfo{author}{\bibfnamefont{S.}~\bibnamefont{Lepri}},
  \bibinfo{author}{\bibfnamefont{C.}~\bibnamefont{Mej{\'\i}a-Monasterio}},
  \bibnamefont{and} \bibinfo{author}{\bibfnamefont{A.}~\bibnamefont{Politi}},
  \bibinfo{journal}{J. Phys. A: Math. Theor.} \textbf{\bibinfo{volume}{42}},
  \bibinfo{pages}{025001} (\bibinfo{year}{2009}).

\end{thebibliography}
\bibliographystyle{apsrev}

\end{document}